\documentclass[12pt]{iopart}
\usepackage[T1]{fontenc}
\usepackage[utf8]{inputenc}
\usepackage{iopams}
\usepackage{graphicx}
\usepackage{url}
\usepackage{cleveref} % for cref
\usepackage{color}
\usepackage{braket}
\usepackage[dvipsnames]{xcolor}
\usepackage{fancyhdr}

\newcommand{\enm}{E^{\rm NM}}
\newcommand{\rhosat}{\rho_{\rm sat}}
\newcommand{\knm}{K^{\rm NM}}
\newcommand{\asym}{a_{\rm sym}^{\rm NM}}
\newcommand{\lsym}{L_{\rm sym}^{\rm NM}}

\newcommand{\msca}{m_s^{*}}
\newcommand{\mvec}{m_v^{*}}
\newcommand{\Crr}[1]{C_{#1}^{\rho\rho}}
\newcommand{\Crt}[1]{C_{#1}^{\rho\tau}}
\newcommand{\CrDr}[1]{C_{#1}^{\rho\Delta\rho}}
\newcommand{\CrDJ}[1]{C_{#1}^{\rho\nabla J}}
\newcommand{\CJJ}[1]{C_{#1}^{JJ}}
\newcommand{\Cuu}[1]{C_{#1}^{uu'}}
\newcommand{\VZeroN}{V_0^n}
\newcommand{\VZeroP}{V_0^p}
\newcommand{\VZeroq}{V_0^q}
\newcommand{\rvec}{\boldsymbol{r}}

\newcommand{\HFODD}{\textsc{hfodd}}

\newcommand{\UNEDFONE}{\textsc{unedf1}}

\newcommand{\UNEDFONEHFB}{\textsc{unedf1}$_{\rm HFB}$}

\newcommand{\tensor}[1]{\mathsf{#1}}
\newcommand{\gras}[1]{\boldsymbol{#1}}

\newcommand{\beqn}{\begin{eqnarray}}
\newcommand{\eeqn}{\end{eqnarray}}
\newcommand{\be}{\begin{equation}}
\newcommand{\ee}{\end{equation}}
\newcommand{\ba}{\begin{array}}
\newcommand{\ea}{\end{array}}

\newcommand{\DFT}{\textsc{dft}}

\newcommand{\HFB}{\textsc{hfb}}
\newcommand{\EDF}{\textsc{edf}}

\newcommand{\EA}{$E_{\rm A}$}
\newcommand{\EB}{$E_{\rm B}$}
\newcommand{\EFI}{$E_{\rm FI}$}

\newcommand{\Xvec}{\mathbf{X}}
\newcommand{\xb}{\mathbf{x}}
\newcommand{\hyperp}{\gras{\kappa}}
\newcommand{\LHS}{\textsc{lhs}}
\newcommand{\GASP}{\textsc{g}a\textsc{sp}}
\newcommand{\MCMC}{\textsc{mcmc}}
\newcommand{\MAP}{\textsc{map}}
\newcommand{\ErrEmul}{\epsilon_{\rm GaSP}}

\begin{document}

\title[A Bayesian Analysis of Nuclear Deformation Properties]{A Bayesian Analysis of Nuclear Deformation Properties with Skyrme Energy Functionals}

\author{N. Schunck$^1$, K. R. Quinlan$^2$, J. Bernstein$^2$}

\address{$^1$ Nuclear and Chemical Sciences Division, Lawrence Livermore
National Laboratory, Livermore, CA 94551, USA}
\address{$^2$ Applied Statistics Group, Lawrence Livermore National Laboratory, Livermore, CA 94551, USA}

\ead{schunck1@llnl.gov}

\vspace{10pt}
\begin{indented}
\item[]May 2020
\end{indented}

\begin{abstract}
In spite of numerous scientific and practical applications, there is still no 
comprehensive theoretical description of the nuclear fission process based 
solely on protons, neutrons and their interactions. The most advanced 
simulations of fission are currently carried out within nuclear density 
functional theory ({\DFT}). In spite of being fully quantum-mechanical and 
rooted in the theory of nuclear forces, {\DFT} still depends on a dozen or so 
parameters characterizing the energy functional. Calibrating these parameters 
on experimental data results in uncertainties that must be quantified for 
applications. This task is very challenging because of the high computational 
cost of {\DFT} calculations for fission. In this paper, we use Gaussian 
processes to build emulators of {\DFT} models in order to quantify and 
propagate statistical uncertainties of theoretical predictions for a range of 
nuclear deformations relevant to describing the fission process.
\end{abstract}


\submitto{\JPG}

\thispagestyle{fancy}

%%%%%%%%%%%%%%%%%%%%%%%%%%%%%%%%%%%%%%%%%%%%%%%%%%%%%%%%%%%%%%%%%%%%%%%%%%%%%%%
%%%%%%%%%%%%%%%%%%%%%%%%%%%%%%%%%%%%%%%%%%%%%%%%%%%%%%%%%%%%%%%%%%%%%%%%%%%%%%%
%%%%%%%%%%%%%%%%%%%%%%%%%%%%%%%%%%%%%%%%%%%%%%%%%%%%%%%%%%%%%%%%%%%%%%%%%%%%%%%
%%%%%%%%%%%%%%%%%%%%%%%%%%%%%%%%%%%%%%%%%%%%%%%%%%%%%%%%%%%%%%%%%%%%%%%%%%%%%%%

\section{Introduction}
\label{sec:introduction}

Nuclear fission plays a key role in a number of basic and applied science 
problems, from understanding the origin of elements in the universe 
\cite{mumpower2016} to the stability of superheavy elements \cite{giuliani2019}. 
From a fundamental science perspective, one would like to be able to describe 
nuclear fission in a fully quantum-mechanical way as emerging from nuclear 
forces within the nucleus. Such a ``microscopic'' picture poses formidable 
challenges to theorists as it would require solving the quantum many-body 
problem of fermions in interaction even though nuclear forces remain poorly 
known. Currently, nuclear density functional theory represents our best 
attempt to tackle such a problem \cite{schunck2016}.

Density functional theory ({\DFT}) is a general approach to the quantum 
many-body problem that is designed to scale well with particle number 
\cite{dreizler1990,eschrig1996}. In the particular case of nuclear physics, 
effective, in-medium nuclear forces are encoded in an energy density 
functional {({\EDF}), which is a function of the intrinsic density of nucleons. 
Given this {\EDF}, various theoretical techniques allow computing a number of 
nuclear properties ranging from ground-state properties, low-lying excited 
spectra or large-amplitude collective motion such as fission or nuclear 
reactions \cite{schunck2019}. 

While {\DFT} has been successfully applied in many areas of nuclear science, 
it should be viewed as an imperfect, phenomenological model. In particular, 
the parameters of the {\EDF} are unknown and must be calibrated on a set of 
experimental data, and this process depends on the particular level of 
approximation within {\DFT} \cite{schunck2015,schunck2015a,schunck2015c}. 
Over the past decade, there have been numerous attempts to quantify the 
uncertainties of this calibration on predictions, but these earlier studies 
have focused mostly on various ground-state properties such as masses 
\cite{gao2013,goriely2014,utama2016,haverinen2017}, drip-line properties 
\cite{neufcourt2018,neufcourt2019}, or neutron skin \cite{kortelainen2013,
reinhard2013}. There are still relatively few examples where either covariance 
or Bayesian techniques were applied to more complex problems such as collective 
excitations \cite{paar2014} or fission barriers \cite{mcdonnell2015}. Yet, in 
light of the computational cost of fission calculations, such analyses are 
essential to better identify model weaknesses.

The main goal of this paper is to assess whether standard statistical methods 
such as Bayesian inference with Gaussian processes can be used with 
confidence in the emulation and uncertainty quantification of fission models. 
Specifically, we wish to extend the work of \cite{mcdonnell2015} to (i) build 
an emulator of fission pathways and (ii) build an emulator of the location and 
characteristics of scission configurations (the point where the two fragment 
are formed) and (iii) determine the posterior distribution of {\EDF} parameters 
conditioned on fission barriers .

Our paper is organized in three main sections. \Sref{sec:physics} recalls some 
basic elements of nuclear energy density functional theory for the particular 
case of Skyrme functionals. In \sref{sec:stat}, we summarize the statistical 
models that we used in the analysis including Gaussian processes and Bayesian 
inference with Markov-Chain Monte-Carlo sampling. Finally, we show in 
\sref{sec:results} a selection of results for the benchmark case of the 
potential energy curve of the $^{240}$Pu nucleus.

%%%%%%%%%%%%%%%%%%%%%%%%%%%%%%%%%%%%%%%%%%%%%%%%%%%%%%%%%%%%%%%%%%%%%%%%%%%%%%%
%%%%%%%%%%%%%%%%%%%%%%%%%%%%%%%%%%%%%%%%%%%%%%%%%%%%%%%%%%%%%%%%%%%%%%%%%%%%%%%
%%%%%%%%%%%%%%%%%%%%%%%%%%%%%%%%%%%%%%%%%%%%%%%%%%%%%%%%%%%%%%%%%%%%%%%%%%%%%%%
%%%%%%%%%%%%%%%%%%%%%%%%%%%%%%%%%%%%%%%%%%%%%%%%%%%%%%%%%%%%%%%%%%%%%%%%%%%%%%%

\section{Physics Background}
\label{sec:physics}

In this section, we give a brief summary of the (single-reference) energy 
density functional theory with Skyrme generators that will be used throughout
this paper. 

%%%%%%%%%%%%%%%%%%%%%%%%%%%%%%%%%%%%%%%%%%%%%%%%%%%%%%%%%%%%%%%%%%%%%%%%%%%%%%%
%%%%%%%%%%%%%%%%%%%%%%%%%%%%%%%%%%%%%%%%%%%%%%%%%%%%%%%%%%%%%%%%%%%%%%%%%%%%%%%

\subsection{Nuclear Energy Density Functional Theory}
\label{subsec:theory}

The general framework for this work is the Hartree-Fock-Bogoliubov ({\HFB}) 
theory, where the nuclear many-body wave function takes the form of a 
quasiparticle vacuum \cite{schunck2019,blaizot1985,ring2004}. In the {\HFB} 
approximation, all degrees of freedom for the system are encoded in the 
one-body density matrix $\rho$ and pairing tensor $\kappa$ (and its complex 
conjugate $\kappa^{*}$). In particular, the total energy of the nucleus reads
\be
E[\rho,\kappa,\kappa^{*}] = E_{\rm nuc}[\rho] + E_{\rm Cou}[\rho] + 
E_{\rm pair}[\rho,\kappa,\kappa^{*}] \,.
\label{eq:etot}
\ee
Here, we use a Skyrme-like energy density functional ({\EDF}) for the nuclear 
part, 
\be
E_{\rm nuc}[\rho] = \sum_{t=0,1} \int d^{3}\gras{r}\; \chi_t(\gras{r}) \,,
\ee
where the Skyrme {\EDF} includes the kinetic energy term and reads
\be
\chi_t(\gras{r}) = \Crr{t} \rho_t^2 + \Crt{t} \rho_t\tau_t 
+ \CJJ{t} \tensor{J}^{2}_t + \CrDr{t}\rho_t\Delta\rho_t 
+ \CrDJ{t}\rho_t\gras{\nabla}\cdot\gras{J}_t \,.
\ee
In this expression, the index $t$ refers to the isoscalar ($t=0$) or isovector 
$(t=1$) channel. We refer to Refs.~\cite{engel1975,dobaczewski1996,bender2003,
perlinska2004,lesinski2007} for the definition of the densities $\rho$, $\tau$,  
$\tensor{J}$, and $\gras{J}$. There are 8 real-valued coupling constants 
$\Cuu{t}$ for $t=0,1$ and $uu' = (\rho\tau, JJ, \rho\Delta\rho, \rho\nabla J)$. 
The case of $\Crr{t}$ is a little different, since it is a function of the 
density $\rho(\rvec)$, 
\be
\Crr{t} = \Crr{t0} + \Crr{tD} \rho^{\gamma}_{0}(\rvec) \, .
\ee
This density-dependent term is thus characterized by 5 parameters, the two 
$C_{t0}^{\rho\rho}$ and $C_{t{\rm D}}^{\rho\rho}$ and the exponent $\gamma$. 
Therefore, $E_{\rm nuc}[\rho]$ is fully characterized by 13 parameters.

The Coulomb term in \eref{eq:etot} is computed at the Hartree-Fock 
approximation with the exchange term treated at the Slater approximation 
\cite{bender2003}. The pairing energy is computed from a surface-volume 
density-dependent pairing force 
\be
V_{q}(\rvec,\rvec') = 
\VZeroq
\left[ 1 - \frac{1}{2}\frac{\rho(\rvec)}{\rho_{c}} \right ]
\delta(\rvec-\rvec') \,,
\ee
where $q$ here refers to the type of particle (proton or neutron) and $\rho_{c} 
= 0.16 $ fm$^{-3}$. Including the pairing channel in the fit thus adds 2 more 
parameters to the fit. As customary for zero-range pairing forces, a cut-off at 
$E_{\rm cut} = 60$ MeV is introduced to limit the number of quasiparticles used 
when calculating the density matrix.

In the context of fission, potential energy curves (or surfaces) are obtained 
by adding constraints on the density $\rho$. The constraints are typically the 
expectation value of suitable operators on the {\HFB} vacuum. In this work, we 
will consider only one constraint on the expectation value of the axial 
quadrupole moment $q \equiv \braket{\hat{Q}_{20}}$. The total {\HFB} energy at 
the deformation $q$ is thus the scalar function $E(q)$. It implicitly depends 
on the vector of coupling constants $\xb \equiv \{ \Cuu{t} \}$ of the energy 
density functional, $E \equiv E(q; \xb)$.

%%%%%%%%%%%%%%%%%%%%%%%%%%%%%%%%%%%%%%%%%%%%%%%%%%%%%%%%%%%%%%%%%%%%%%%%%%%%%%%
%%%%%%%%%%%%%%%%%%%%%%%%%%%%%%%%%%%%%%%%%%%%%%%%%%%%%%%%%%%%%%%%%%%%%%%%%%%%%%%

\subsection{Parameters and Numerical Implementation}
\label{subsec:params}

Following \cite{kortelainen2010}, we express the coupling constants 
$\Crr{t0}$, $\Crr{tD}$, $\gamma$ and $\Crt{t}$ as a function of the parameters 
of infinite nuclear matter. Our analysis will be focused on the {\UNEDFONEHFB} 
functional, for which the two tensor coupling constants are 0 and the vector 
effective mass is set at the SLy4 value, $\mvec = 1.249 838$ 
\cite{chabanat1998}. As a result, the actual vector $\xb$ of parameters is
\be
\xb = \left( 
\enm, \rhosat, \knm, \asym, \lsym, \msca,  
\CrDr{0}, \CrDr{1}, \CrDJ{0}, \CrDJ{1}, \VZeroN, \VZeroP
\right)
\ee
and the total number of parameters under consideration is $n=12$.

Calculations were performed with the latest version of the code {\HFODD} 
\cite{schunck2017}. The solutions to the {\HFB} equation are expanded on a 
deformed basis of $N_{\rm shell} = 30$ shells. For each value of the quadrupole 
moment, the oscillator length and axial quadrupole deformation of the basis, 
which determines the ratios of oscillator frequencies $\omega_{z}/\omega_{x}$, 
are set according to the empirical formula of \cite{schunck2014}. The 
Gauss-Hermite integration mesh comprises 40 points in the $x$- and 
$y$-directions and 66 in the $z$-direction. We use linear constraints on the 
quadrupole moment (and on the dipole moment to fix the position of the center 
of mass), and the value of the Lagrange parameter is set at each iteration 
based on the cranking approximation of the \textsc{qrpa} matrix.

%%%%%%%%%%%%%%%%%%%%%%%%%%%%%%%%%%%%%%%%%%%%%%%%%%%%%%%%%%%%%%%%%%%%%%%%%%%%%%%
%%%%%%%%%%%%%%%%%%%%%%%%%%%%%%%%%%%%%%%%%%%%%%%%%%%%%%%%%%%%%%%%%%%%%%%%%%%%%%%
%%%%%%%%%%%%%%%%%%%%%%%%%%%%%%%%%%%%%%%%%%%%%%%%%%%%%%%%%%%%%%%%%%%%%%%%%%%%%%%
%%%%%%%%%%%%%%%%%%%%%%%%%%%%%%%%%%%%%%%%%%%%%%%%%%%%%%%%%%%%%%%%%%%%%%%%%%%%%%%

\section{Statistical methods}
\label{sec:stat}

%%%%%%%%%%%%%%%%%%%%%%%%%%%%%%%%%%%%%%%%%%%%%%%%%%%%%%%%%%%%%%%%%%%%%%%%%%%%%%%
%%%%%%%%%%%%%%%%%%%%%%%%%%%%%%%%%%%%%%%%%%%%%%%%%%%%%%%%%%%%%%%%%%%%%%%%%%%%%%%

In this section we first introduce our Gaussian Process emulator of the 
potential energy curves $E(q; \xb)$ and then describe how this emulator will 
be used to generate a posterior distribution of the parameters. 

%%%%%%%%%%%%%%%%%%%%%%%%%%%%%%%%%%%%%%%%%%%%%%%%%%%%%%%%%%%%%%%%%%%%%%%%%%%%%%%
%%%%%%%%%%%%%%%%%%%%%%%%%%%%%%%%%%%%%%%%%%%%%%%%%%%%%%%%%%%%%%%%%%%%%%%%%%%%%%%

\subsection{Gausian Process Emulator}
\label{subsec:gasp}

To emulate the potential energy curves, we use a Gaussian Stochastic Process 
({\GASP}) emulator as described in \cite{gu2018robust}. To implement a {\GASP}, 
the data should be observed at the same input locations for all curves, i.e., 
all potential energy curves must be defined on the same grid of deformations 
$q$. A {\GASP} $y(\cdot)$ can be defined as
\be
y(\cdot) \sim \mathrm{\GASP}( \mu(\cdot), \sigma^2 C(\cdot, \cdot) ) \,, 
\ee
with $\mu$ the mean function, $\sigma^2$ the variance and $C$ the correlation 
function. Here the mean function is modeled as a linear combination of basis 
functions $\mathbf{h}(\xb) \equiv (h_{1}(\xb),\dots, h_{q}(\xb) )$ and 
regression parameters $\hyperp \equiv (\kappa_{1}, \dots, \kappa_{q})$ so that
\be
\mathbf{\mu}(\xb) = E[y(\xb)] = \mathbf{h}(\xb)\cdot\hyperp 
= \sum_{t=1}^{q} h_{t}(\xb)\kappa_{t} \,.
\ee
The correlation function is defined by the Mat\'ern covariance structure. We 
apply a common specific case referred to as the Mat\'ern 5/2, 
\be
C(\xb,\xb') = 
\left(1 + \frac{\sqrt{5}\ || \xb - \xb'|| }{\gamma} 
+ 
\frac{5 || \xb - \xb'||^2}{3 \gamma^2} \right) 
\exp \left( - \frac{ \sqrt{5} || \xb - \xb'|| }{ \gamma } \right) \,,
\ee
where $\gamma$ is a hyper-parameter that is optimized to obtain a better fit. 
This correlation structure is stationary, meaning that the correlation is 
assumed to only depend on the distance between two observations, i.e., 
$||\xb - \xb'||$, but not based on the location of those observations. Note 
that each deformation $q$ is treated independently of the others. 
This model is fit using the RobustGaSP R package \cite{GaSPpackage}. The 
default setting uses a constant mean for $\mu(\xb)$ but first subtracts the 
mean at each input location. The model is robust in terms of the estimation of 
hyperparameters such as $\gamma$. It avoids numerical issues while still 
achieving good predictive performance.

%%%%%%%%%%%%%%%%%%%%%%%%%%%%%%%%%%%%%%%%%%%%%%%%%%%%%%%%%%%%%%%%%%%%%%%%%%%%%%%
%%%%%%%%%%%%%%%%%%%%%%%%%%%%%%%%%%%%%%%%%%%%%%%%%%%%%%%%%%%%%%%%%%%%%%%%%%%%%%%

\subsection{Bayesian Regression}
\label{subsec:bayes}

We now discuss the Bayesian framework that will be used to find posterior 
distributions of the parameters. The philosophy of Bayesian inference starts 
with the fact that there is a prior probability distribution representing the 
\emph{a priori} beliefs or knowledge about the parameter set. This prior state 
of knowledge is updated when presented with the observed experimental data to 
give a posterior probability distribution representing the current state of 
knowledge given the data. Formally, our goal is to determine the posterior 
distribution of the parameters $\xb$ represented as $\pi(\xb | y)$ using 
observations $y_1,\dots, y_n$ using the formula  
\be
\pi(\xb|y) \propto L_y(\xb) \pi(\xb) \,,
\ee
where $\pi(\xb)$ is the prior distribution and $L$ is the likelihood function 
of the data (the theoretical prediction of the model) given the parameters. 
Here we use a use a normal likelihood function justified by assuming Gaussian measurement errors. Given a set $\{ d_i \}_{i=1,n_{d}}$ 
of $n_{d}$ experimental values (in our case the three measurements in Table \ref{tab:exp_barriers}) for some observables and 
$\{ y_i(\xb) \}_{i=1,n_{d}}$ the theoretical predictions obtained from the 
{\GASP} emulator at $\xb$ for these same observables, then the likelihood 
function reads
\be
L_y(\xb) \propto 
 \left( \prod_{i=1}^{n_{d}} \sigma_{i}^2 \right)^{ \frac{n_d}{2}} \exp\left( -  \sum_{i = 1}^{n_{d}} \frac{(y_i(\xb) - d_i)^2}{2 \sigma_{i}^2} \right)   \,,
\label{func:likelihood}
\ee
where $\sigma^2_i$ is the variance for each experimental value.  
The advantage of Bayesian inference is that our result is a full distribution 
which allows us to make statements about the uncertainty of our parameter 
estimates. The difficulty here is that an analytic form of the posterior 
distribution can only be found under specific distributional assumptions on 
the prior and likelihood function. These assumptions are not always realistic, 
and as such approximations of the posterior are typically made using Markov 
Chain Monte Carlo ({\MCMC}) sampling techniques. These methods yield a large 
number of approximately independent samples from a Markov chain that should 
have the same distribution as the posterior. Here, we select non-informative 
prior distributions that are uniform over the given parameter boundaries. This 
implies that we have no prior knowledge of which region of the parameter space 
is closer to the ``real value''; we only assume that it will occur somewhere 
in the specified region.

We apply a Delayed Rejection Adaptive Metropolis {\MCMC} algorithm 
\cite{haario2006}. The Delayed Rejection algorithm samples a second value when 
a proposed value is rejected. This second proposal is typically drawn from a 
proposal distribution with a smaller variance. We use a maximum number of 
delayed rejections of one at each iteration. It is possible to do this multiple 
times, but it will greatly increase the computation time of the {\MCMC} 
algorithm if the maximum number of delayed rejections at each iteration is too 
large. Delayed rejection is advantageous because it has been shown to give 
smaller asymptotic variances of the estimators from the chain 
\cite{tierney1999}.  The Adaptive Metropolis component updates the covariance 
of the proposal distribution based on previous iterations of the chain. This 
adapts the shape and size of the sampling distribution and generally makes the 
{\MCMC} more efficient. In our implementation this update is only done during 
the burn-in phase.   

%%%%%%%%%%%%%%%%%%%%%%%%%%%%%%%%%%%%%%%%%%%%%%%%%%%%%%%%%%%%%%%%%%%%%%%%%%%%%%%
%%%%%%%%%%%%%%%%%%%%%%%%%%%%%%%%%%%%%%%%%%%%%%%%%%%%%%%%%%%%%%%%%%%%%%%%%%%%%%%
%%%%%%%%%%%%%%%%%%%%%%%%%%%%%%%%%%%%%%%%%%%%%%%%%%%%%%%%%%%%%%%%%%%%%%%%%%%%%%%
%%%%%%%%%%%%%%%%%%%%%%%%%%%%%%%%%%%%%%%%%%%%%%%%%%%%%%%%%%%%%%%%%%%%%%%%%%%%%%%

\section{Results}
\label{sec:results}

In this section, we focus on the one-dimensional potential energy curve of 
$^{240}$Pu as as function of the axial quadrupole moment, from the ground state 
to the point of scission. We first describe the calculations performed to 
generate the training data for the Gaussian process analysis. We then discuss 
the emulator of the potential energy curve before determining the posterior 
distribution of the {\EDF} parameters.

%%%%%%%%%%%%%%%%%%%%%%%%%%%%%%%%%%%%%%%%%%%%%%%%%%%%%%%%%%%%%%%%%%%%%%%%%%%%%%%
%%%%%%%%%%%%%%%%%%%%%%%%%%%%%%%%%%%%%%%%%%%%%%%%%%%%%%%%%%%%%%%%%%%%%%%%%%%%%%%

\subsection{Training runs}
\label{subsec:training}

Gaussian processes must be trained on some input data. In our case, these 
consist of a set of potential energy curves for a set of $N$ parametrizations 
$\Xvec = (\xb_{1},\dots,\xb_{N})$ of the Skyrme functional. For each 
parametrization $\xb_{k}$, we must calculate the full potential energy curve -- 
here with triaxial deformations included. Since the parameter space of 
{\UNEDFONEHFB} is 12-dimensional, the amount of calculation needed could 
quickly become gigantic. We thus imposed two restrictions to alleviate the 
computational cost:
\begin{itemize}
\item We reduced the number of varying coupling constants to 6 by keeping all 
bulk coupling constants fixed at their {\UNEDFONEHFB} value. This is justified 
since these are related to nuclear matter properties and the result of the 
optimization show that they are relatively well-constrained 
\cite{kortelainen2012,schunck2015a};
\item For each remaining coupling constant, we considered a rather small 
interval of variation centered around the nominal {\UNEDFONEHFB} value, 
$I_{t}^{uu'} = [ \Cuu{t} - \Delta \Cuu{t}, \Cuu{t} + \Delta \Cuu{t}]$. 
The quantities $\Delta \Cuu{t}$ are listed in \tref{tab:intervals} and 
correspond approximately to the standard deviations of the {\UNEDFONE} fit; 
see \cite{kortelainen2012}.
\end{itemize}

\Table{\label{tab:intervals} Range of variation around the {\UNEDFONEHFB} value 
for the six coupling constants included in this work. $\CrDr{t}$ and $\CrDJ{t}$ 
are in MeV fm$^{5}$ and $\VZeroN$ and $\VZeroP$ in MeV fm$^{3}$.}
\br
& $\CrDr{0}$ & $\CrDr{1}$ & $\VZeroN$ & $\VZeroP$ & $\CrDJ{0}$ & $\CrDJ{1}$ \\
\mr
$\Delta \xb $  & 5 & 50 & 20 & 15 & 5 & 25 \\
\br
\end{tabular}
\end{indented}
\end{table}

The product set of all intervals $I_{t}^{uu'}$ defines an hypercube in the 
$6$-dimensional parameter space. We sampled this hypercube with a Latin 
Hypercube Sampling ({\LHS}) algorithm to determine the vector $\Xvec$ of 
parameterizations for the training runs. In practice, as we will show below, 
$N=70$ samples were sufficient to build a high-quality emulator of the model. 
These included 10 runs selected using the Integrated Mean Square Prediction 
Error criterion (IMSPE) \cite{santner2003} to improve estimation of the 
scission point. \Fref{fig:energy} shows the deformation energy (with respect to 
the ground-state value) for each of the 70 samples for the case of $^{240}$Pu. 
The end-point of each curve, usually beyond $q_{20} > 350$ b, represents the 
scission configuration: immediately beyond this point, the total energy drops 
rapidly, which manifests itself by a discontinuity in the energy curve. To 
increase the legibility of the figure, we did not represent this discontinuity 
and simply stopped the curve at scission.

\begin{figure}[!ht]
\begin{center}
\includegraphics[width=0.95\linewidth]{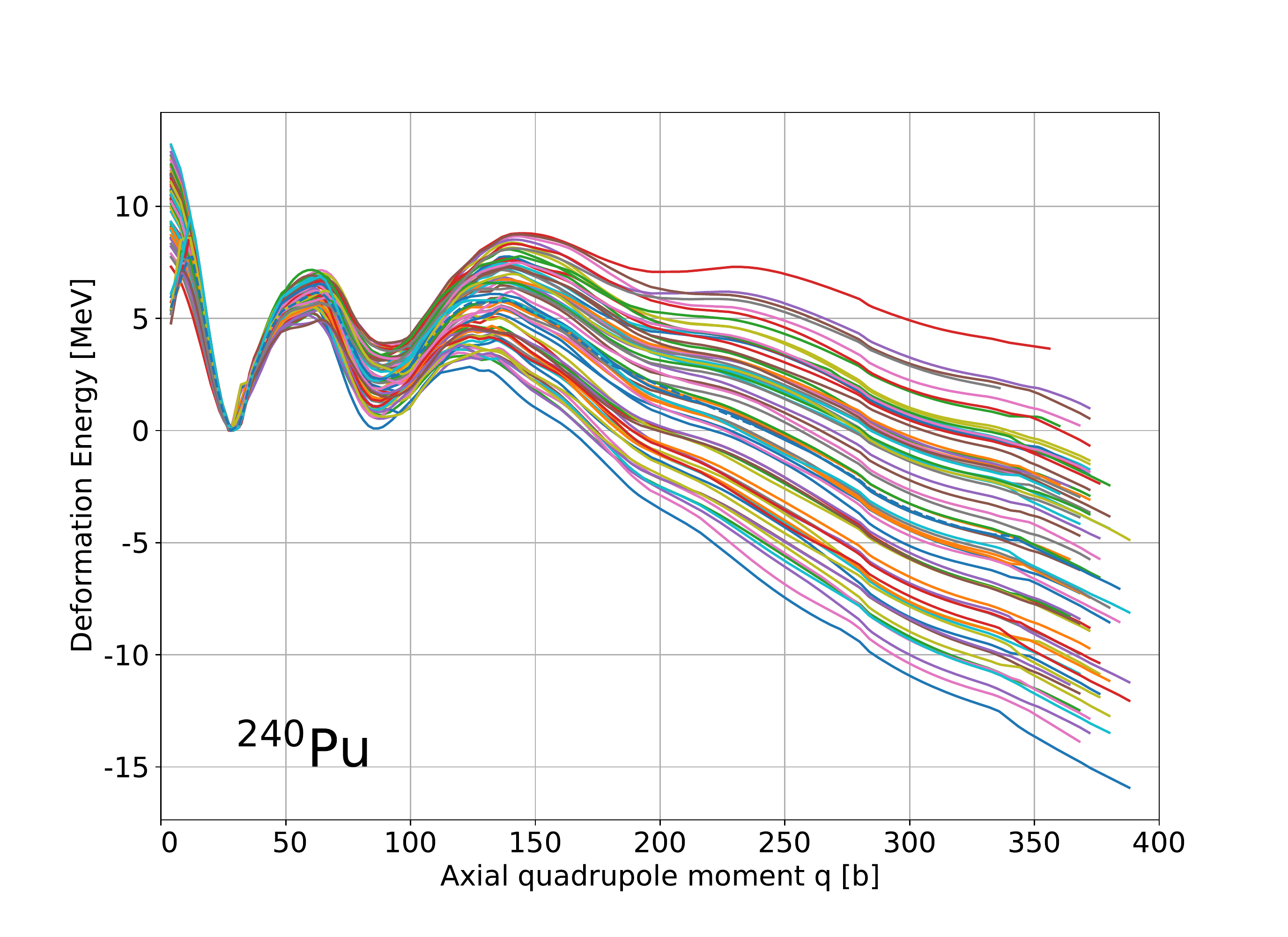}
\caption{Deformation energy curves in $^{240}$Pu as a function of the axial 
quadrupole moment $q\equiv \langle \hat{Q}_{20} \rangle$ for a training set of 
60 different parametrizations of the Skyrme {\EDF}. The rectangular grid guides 
the eye.}
\label{fig:energy}
\end{center}
\end{figure}

%%%%%%%%%%%%%%%%%%%%%%%%%%%%%%%%%%%%%%%%%%%%%%%%%%%%%%%%%%%%%%%%%%%%%%%%%%%%%%%
%%%%%%%%%%%%%%%%%%%%%%%%%%%%%%%%%%%%%%%%%%%%%%%%%%%%%%%%%%%%%%%%%%%%%%%%%%%%%%%

\subsection{Emulator Fit}

In this section we fit the {\GASP} model to the potential energy curves. For 
the dataset of 70 training runs, the observations were not equally spaced, but 
they were dense. Thus, the points used to fit the model were selected by equal 
spacing via interpolation. Since the scission point was not needed for 
calibration, the emulator was only run up to $q = 300$ b. We are not only 
interested in having accurate estimation of the potential energy curve for the 
parameter sets we have observed, but also for any parameter set within the 
bounds of interest. Thus, we will use cross validation to look at the out of 
sample error for our emulator, specifically applying leave-one-out cross 
validation. This is effective because it does not require additional samples, 
and shows what the model predictions would have been for each parameter set had 
the potential energy not been observed. \Fref{fig:GaSPLOO} gives the 
leave-one-out residuals for all 70 curves. To obtain leave-one-out residuals 
for curve $y_i$, we first fit the model using the other 69 observations. Then, 
we use the predicted mean curve $\hat{y}_{i}$ using the parameter set $\xb_{i}$ 
to obtain the residual. 

\begin{figure}[!ht]
\begin{center}
\includegraphics[width=0.85\linewidth]{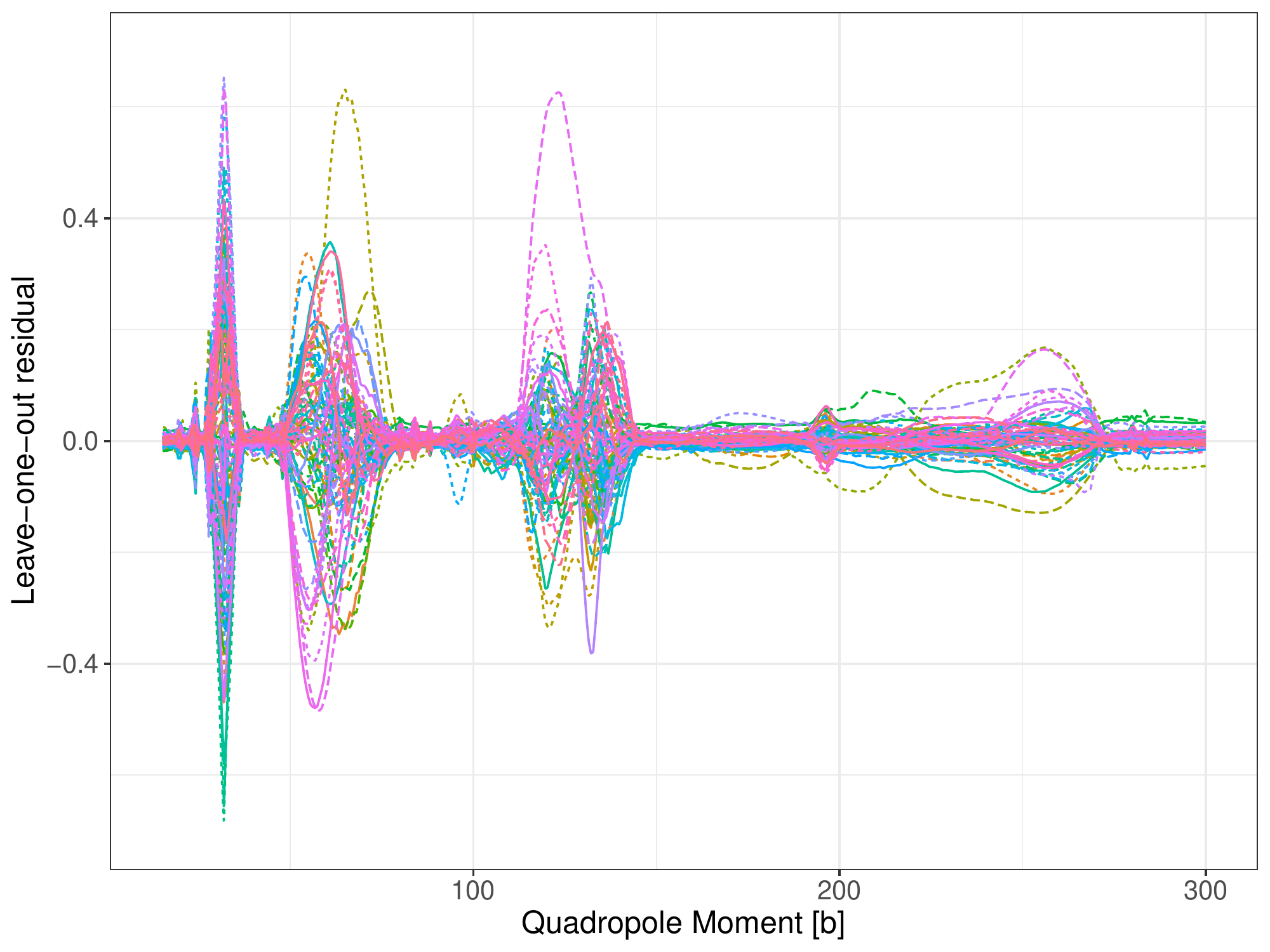}
\caption{Leave-one-out residuals $\ErrEmul = E(q) - E_{\rm GaSP}(q)$ (in MeV) 
for the {\GASP} emulator as a function of the axial quadrupole moment.}
\label{fig:GaSPLOO}
\end{center}
\end{figure}

The results show that the largest errors $\ErrEmul$ occur around the local 
maxima/minima of the potential energy curve with the areas in between having 
much smaller residuals. In fact, a closer analysis of the emulation error shows 
that it is directly connected with the onset of triaxiality. This is better 
visualized in \fref{fig:error}, which shows the expectation values 
$\braket{\hat{Q}_{22}}$ and $\braket{\hat{Q}_{30}}$ as a function of the axial 
quadrupole moment. As a reminder, the degree of triaxiality of the nuclear 
shape is typically characterized by the Bohr $\gamma$ angle. With the 
conventions adopted for the multipole moments in {\HFODD}, we have 
$\gamma = \mathrm{atan}\big(\braket{\hat{Q}_{22}}/\braket{\hat{Q}_{20}}\big)$. 
Therefore, the shape is prolate axial if $\gamma = 0^{\mathrm{o}}$ (hence 
$\braket{\hat{Q}_{22}}=0$ b), oblate axial if $\gamma = 60^{\mathrm{o}}$ 
($\braket{\hat{Q}_{22}} = \sqrt{3}\braket{\hat{Q}_{20}}$), and it is 
maximally triaxial if $\gamma = 30^{\mathrm{o}}$ ($\braket{\hat{Q}_{22}} = 
\sqrt{1/3}\braket{\hat{Q}_{20}}$). 

The onset of triaxiality at $q \approx 35$ b (depending on the parameter set), 
the return to axial symmetry at $q \approx 60$ b, and the slightly trixial 
shapes between $110\ \mathrm{b}\geq q \geq 145$ b correspond exactly to the 
regions where the emulator error is maximal, reaching up to 500 keV. This is 
most likely the consequence of spurious discontinuities that appear in the 
self-consistent calculations when projecting on one-dimensional paths as in 
\fref{fig:GaSPLOO}. Indeed, while the expectation value of the axial 
quadrupole moment $q$ is constrained, every other moment of the nuclear surface 
is unconstrained. The variational principle dictates that the {\HFB} 
calculation will converge to the \emph{local} minimum nearest to the starting 
point. In the higher-dimensional space characterized by an additional 
collective variable $q'$, one may find at point $q = q_{0}$ two minima 
quasi-degenerate in energy but separated by a barrier along the $q'$ direction: 
projecting such a surface on the $q$ axis would give a continuous, 
non-differentiable curve at $q_{0}$; see discussion in \cite{dubray2012}. This 
is exactly what seems to happen here, with $q' \equiv q_{22}$ controlling the 
degree of triaxiality. By contrast, the onset of octupole deformation around 
$q \approx 100$ b does not produce any noticeable increase in emulation error.

\begin{figure}[!ht]
\begin{center}
\includegraphics[width=0.95\linewidth]{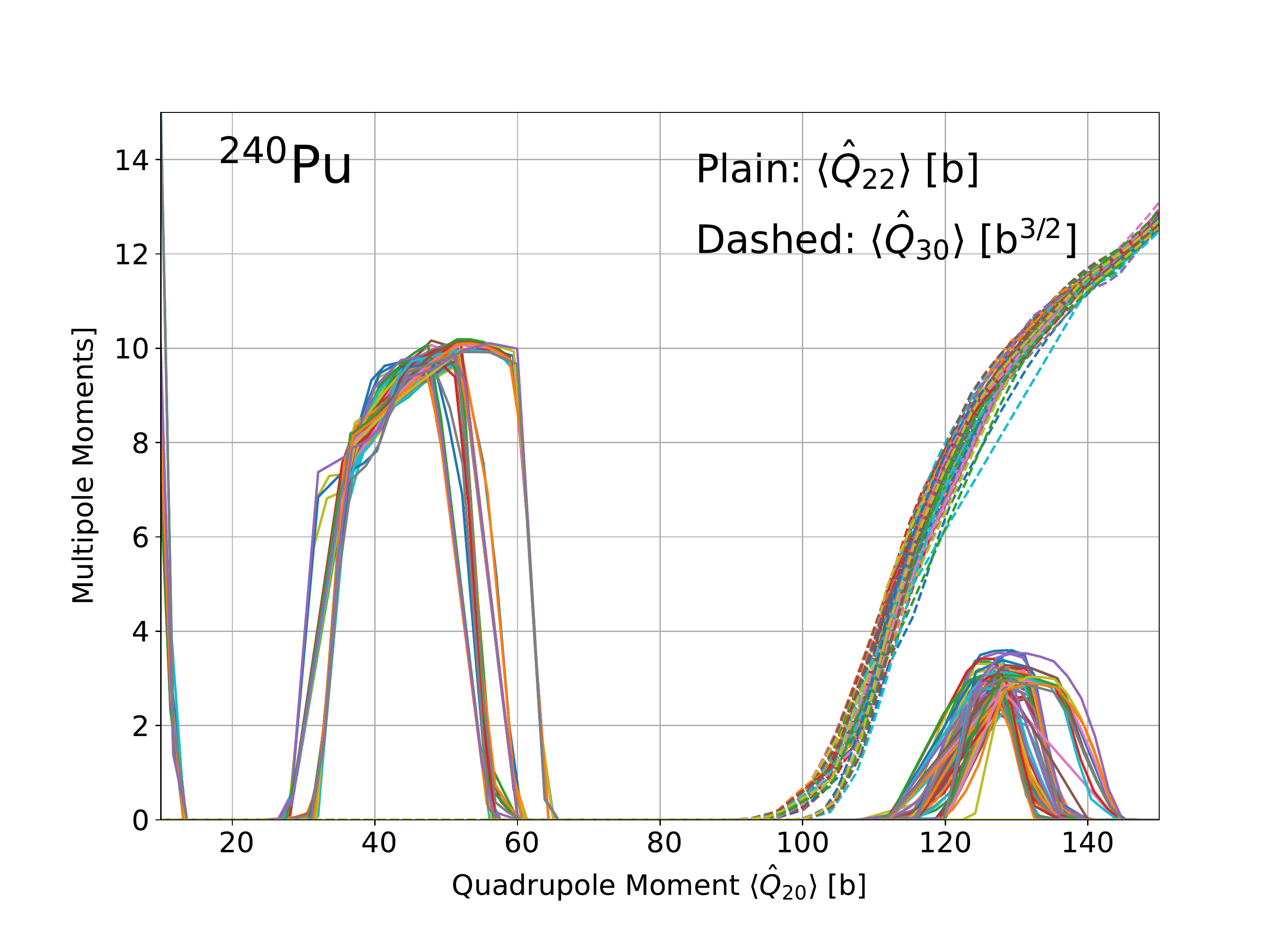}
\caption{Expectation value $\braket{\hat{Q}_{22}}$ (triaxial quadrupole) and 
$\braket{\hat{Q}_{30}}$ (axial octupole) of the multipole moments as a function 
of the axial quadrupole moment $q\equiv \braket{\hat{Q}_{20}}$.}
\label{fig:error}
\end{center}
\end{figure}

The size of this emulation error $|\ErrEmul| \leq 0.68$ MeV should be compared 
to other estimates of relevant uncertainty: (i) for triaxial {\HFB} 
calculations based on expansions in the one-center harmonic oscillator basis, 
basis truncation errors alone can easily reach a few MeV \cite{schunck2013} 
(ii) while the experimental value of the fission isomer excitation energy in 
$^{240}$Pu is known to within $\pm 0.2$ MeV \cite{hunyadi2001}, this number 
comes from a single experiment and older estimates differ by 0.6 MeV 
\cite{singh2002} (iii) the height of both fission barriers is a model-dependent 
quantity extracted from fission cross sections, and the uncertainty is 
typically of the order of 1 MeV; see discussion in \sref{subsec:posterior} 
below.

\Fref{fig:centeredruns} shows four examples of the centered potential energy 
curves along with the leave-one-out mean prediction and uncertainty from the 
{\GASP} emulator. Given the $n=70$ samples, the centered potential energy curve 
$E(q,\xb_{i})$ is obtained by subtracting at each point the mean value across 
all samples,
\be
E_{\rm cent}(q,\xb_{i}) = E(q,\xb_{i}) - \frac{1}{n}\sum_{k=1}^{n} E(q,\xb_{k}).
\ee
Centered potential energy curves with mostly negative values simply indicate 
that the original potential energy at each deformation $q$ is lower than the 
average value at that deformation over the $n$ samples. The four parameter sets 
were chosen for visual clarity and are representative of the typical 
performance of the emulator. From this we can see that the emulator is highly 
confident and also accurately captures the potential energy curves. There is 
more uncertainty at the minima and maxima of the curve, so the larger observed 
deviations from the mean prediction still largely fall within the 95\% credible 
intervals.

\begin{figure}[!ht]
\begin{center}
\includegraphics[width=0.9\linewidth]{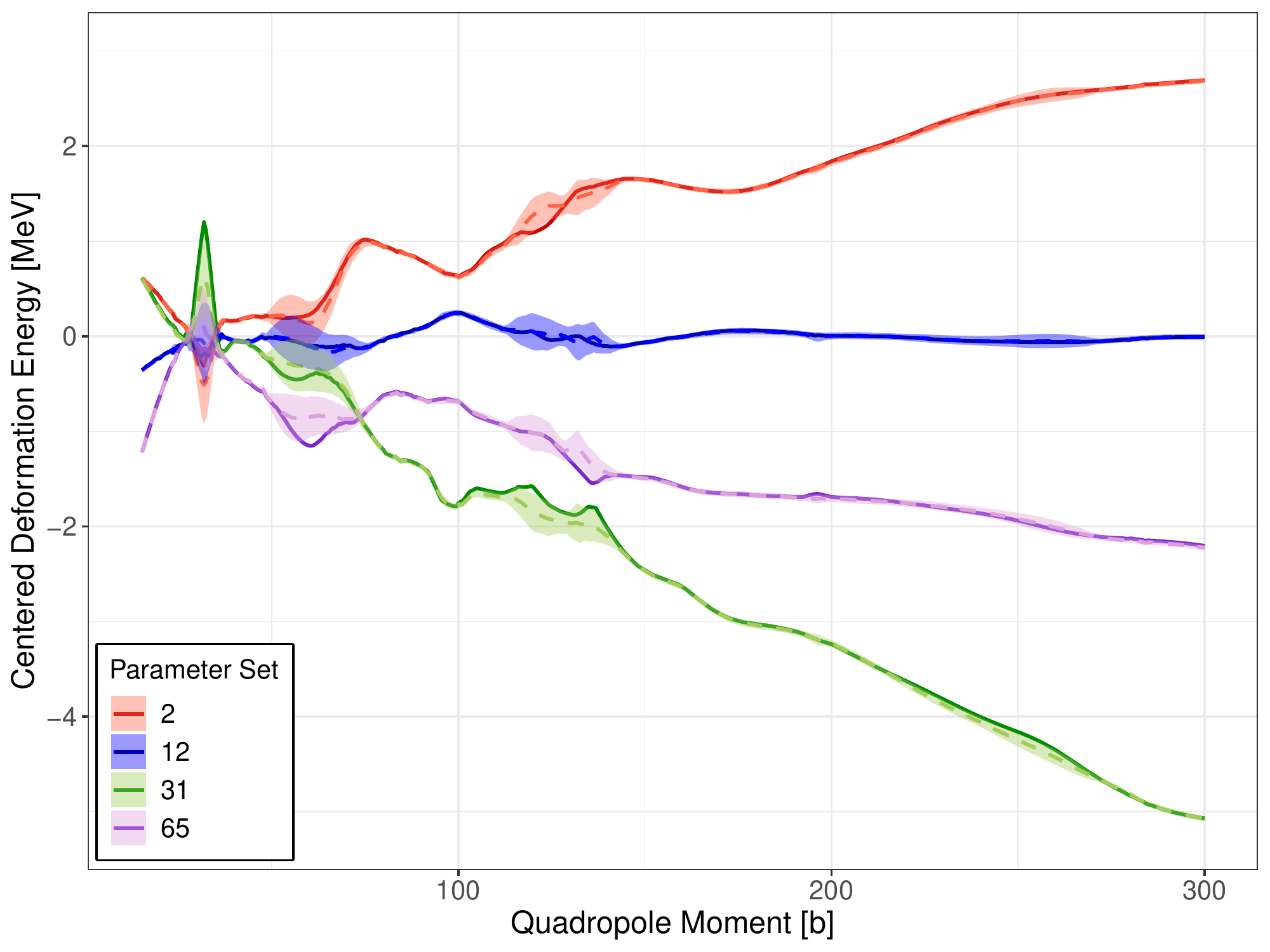}
\caption{Centered potential energy curves with {\GASP} predictions. Dashed lines 
represent the mean prediction of the {\GASP} model, and shaded areas represent 
95\% credible intervals. The solid lines represent the simulated values. }
\label{fig:centeredruns}
\end{center}
\end{figure}

Many relevant fission observables cannot be computed with a single collective 
variable $q$ but require a higher-dimensional collective space. This is the 
case, for instance, with the distributions in charge, mass and kinetic energy 
of the fission fragments. Most importantly, these observables also need to be 
computed at scission, i.e., just before the system has split into two 
fragments. As mentioned in the previous section, in a one-dimensional space, 
scission corresponds to a point (the end-point of each curve in 
\fref{fig:energy}); in a collective space of dimensions $D$, scission 
configurations correspond to a $(D-1)$-dimensional hypersurface. The 
characteristics of this surface are a function of the parameters $\xb$ of the 
functional which we need to emulate. Before embarking in such a project, which 
would imply running expensive calculations in a $D$-dimensional space, it is 
worth checking the ability of Gaussian processes to reproduce the scission 
point already in $D=1$. In practice, we try to emulate the location 
$q_{\rm scis}$ of the discontinuity in the potential energy curve as a function 
of $\xb$, as well as the number of particles $Z_{H}$ and $N_{H}$ of the heavy 
fragment at that point. These values are scalars and we use a simple Gaussian 
Process for prediction. 

The leave-one-out residuals are shown in \fref{fig:boxplot} in the form of 
boxplots and ``violin'' plots. The upper and lower portions of the box portion 
of the boxplot show the 25th and 75th percentile of the data, and the 
horizontal line in the middle shows the median. The minimum and maximum values 
are indicated by the ends of the vertical line going through the box; except 
for cases where values are far enough away form the box portion of the data 
(traditionally found as 1.5 $\times$ (75th percentile - 25th percentile) above 
or below the box portion). In these cases the individual points are plotted, 
such as the largest negative residual for the scission point. The shaded area 
is the violin plot and it gives an estimate of the probability distribution of 
the data.  

\begin{figure}[!ht]
\begin{center}
\includegraphics[width=0.9\linewidth]{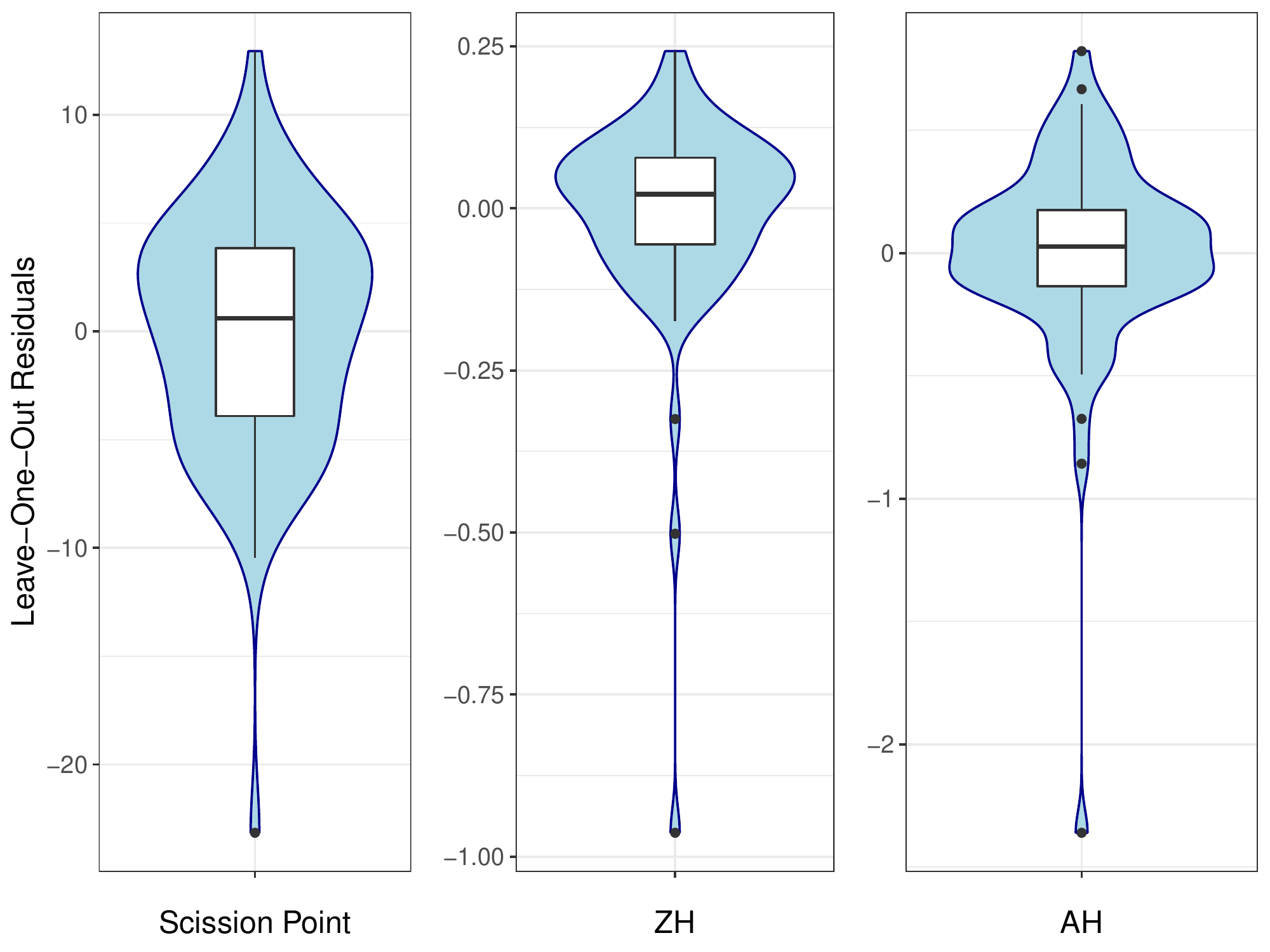}
\caption{Boxplots overlaying violin plots for the leave-one-out prediction 
error for the location of the scission point, the number of protons in the 
heavy fragment (ZH) and the number of particles in the heavy fragment (AH)}
\label{fig:boxplot}
\end{center}
\end{figure}

Disregarding the outliers for the moment, we see that on average, the location 
of the scission point is reproduced within about 5 -- 10 barns. The impact on 
the number of particles in the heavy fragment is of the order of half a 
particle in total. From a physics perspective, these numbers may look small, 
but one should bear in mind that they were obtained for the case of a 
one-dimensional collective space only: they are likely to increase with the 
size of the collective space and/or the size of the parameter space that we try 
to cover with the emulator even as they decrease with the number of training 
runs. It is not unreasonable to believe that such emulators could lead to 
uncertainties $\sigma_{A} > 1$ in the number of particles at scission. Such 
uncertainties have to be added to the others related to the very concept of 
scission; see discussion in \cite{schunck2016}. It is also interesting to note 
that even though there is only one outlier for the emulation of the scission 
point, there are 3 of them for $Z_{H}$ and $A_{H}$. This could be another 
manifestation of the aforementioned discontinuities, this time at scission. 
This effect is also likely to be further magnified when moving to 
higher-dimensional collective spaces. One may conclude from these results that 
Gaussian processes may not be the best tool to emulate the characteristics of 
scission.

%%%%%%%%%%%%%%%%%%%%%%%%%%%%%%%%%%%%%%%%%%%%%%%%%%%%%%%%%%%%%%%%%%%%%%%%%%%%%%%
%%%%%%%%%%%%%%%%%%%%%%%%%%%%%%%%%%%%%%%%%%%%%%%%%%%%%%%%%%%%%%%%%%%%%%%%%%%%%%%

\subsection{Posterior Parameter Estimates}
\label{subsec:posterior}

With the relative accuracy of the emulator established, we now proceed to 
obtain posterior distributions of the parameters using the {\GASP} emulator by 
conditioning on the experimental value of fission isomer excitation energy and 
fission barriers. This exercise presents two difficulties. First, there is 
little experimental information on the excitation energy of fission isomers 
and, more generally, the band-head of the lowest rotational band built on 
superdeformed minima \cite{singh2002}. Second, fission barriers are extracted 
from fission cross-sections in a model-dependent procedure. Typically, this 
analysis is based on assuming a one-dimensional, inverted parabola for the 
barrier \cite{bjornholm1980}. As a result, fission barriers are not genuine 
experimental data, but empirical one, sometimes called ``metadata''. 
\Tref{tab:exp_barriers} summarizes what is available in the literature for the 
case of $^{240}$Pu.

\Table{\label{tab:exp_barriers} Summary of experimental and empirical 
information about the fission isomer excitation energy of $^{240}$Pu and its 
two fission barriers. The last column is our weighted average. All units are 
MeV. Data for Bjornholm \& Lynn is from \cite{bjornholm1980}; for Capote from 
\cite{capote2009}; for Hilaire from \cite{goriely2009}; for Hunyadi from 
\cite{hunyadi2001} and for Singh from \cite{singh2002}.}
\br
    & Bjornholm & Capote &  Hilaire &    Hunyadi      &     Singh & Weighted \\
\mr
$E_{\rm FI}$  &   2.40 $\pm$ 0.20 &   -    &     -    & 2.25 $\pm$ 0.20 & 2.80 
$\pm$ 0.20 & 2.3500\\
$E_{\rm A}$  &   5.60 $\pm$ 0.20 &  6.05  &   5.89   &     -  &    -  & 5.8975 \\
$E_{\rm B}$  &   5.10 $\pm$ 0.20 &  5.15  &   5.73   &     -  &    -  & 5.2825 \\
\br
\end{tabular}
\end{indented}
\end{table}

To generate the posterior distribution we first need to define our likelihood 
function. When using the normal likelihood \eref{func:likelihood}, posterior 
distributions generated about 10\% of potential energy curves with 
$E_{\rm A} < E_{\rm B}$. Since there is a relative consensus in the physics 
community that $E_{\rm A} > E_{\rm B}$, we modified the likelihood in 
\eref{func:likelihood} to be a truncated normal likelihood: $L(y|\xb)$ is 
set to be extremely small if $E_{\rm A} < E_{\rm B}$ so that the corresponding 
$\xb$ will not be accepted during {\MCMC} iterations. 

The values of $d_i$ for the fission isomer excitation energy and the two 
fission barriers were calculated as a weighted average of the experimental or 
empirical values listed in \tref{tab:exp_barriers} based on subjective 
confidence estimates. Specifically, for $E_{\rm FI}$, the Hunyadi value 
accounted for 60\% of the ``experimental'' value while the Bjornholm and Singh 
numbers accounted for 30\% and 10\%, respectively. For both fission barriers, 
the Capote value accounted for 50\% and the Hilaire and Bjornholm numbers for 
25\% each. The last column of \tref{tab:exp_barriers} gives the actual 
weighted values $d_i$ that we used in the regression. For the calculation of 
the likelihood \eref{func:likelihood}, the variances $\sigma_{i}^2$ were taken 
as $\sigma^2_i = \sqrt{ \sigma^2_{\rm exp} + \sigma^2_{i, \rm \GASP}}$, where 
$\sigma_{\rm exp}$ is the experimental error, see \tref{tab:barriers}, and 
$\sigma^2_{i, \rm \GASP}$, is the sum of the emulator standard deviation at the 
two locations used to calculate each excitation energy energy. 

To generate the posterior samples, the first 10,000 {\MCMC} samples were 
discarded in the burn-in period. The chain ran for a total of 7.5 million 
samples, but was thinned down to 100,000 to ensure the posterior samples were 
uncorrelated. The maximum a posteriori ({\MAP}) estimator $\xb^{*}$ was found 
by taking the parameter set with the highest posterior probability. 
\Tref{tab:MAPcoupling} compares the values of the six coupling constants for the 
{\MAP} estimate and for the original {\UNEDFONEHFB} parametrization. Even 
though the intervals of variation around the {\UNEDFONEHFB} values was 
relatively small as shown in \tref{tab:intervals}, most parameters have changed 
significantly. The two-dimensional bivariate representation of the full 
6-dimensional posterior distribution is shown in \fref{fig:posteriorplot}. 
Clearly, the local fit pins down the value of $\CrDr{0}$ and $\VZeroN$, which 
is compatible with the analysis of \cite{nikolov2011,ryssens2019} suggesting 
that surface properties, which are largely dependent on the interplay between 
$\CrDr{0}$ and $\VZeroN$, are highly correlated with the minima and maxima of 
potential energy curves. Conversely, all other coupling constants end at their 
boundaries, which indicates that they are not well-constrained by deformation 
properties.

\Table{\label{tab:MAPcoupling} Values of the the six coupling constants 
included in this work for the original {\UNEDFONEHFB} parameterization and the 
{\MAP} estimate. $\CrDr{t}$ and $\CrDJ{t}$ are in MeV fm$^{5}$ and $\VZeroN$ 
and $\VZeroP$ in MeV fm$^{3}$.}
\br
& $\CrDr{0}$ & $\CrDr{1}$ & $\VZeroN$ & $\VZeroP$ & $\CrDJ{0}$ & $\CrDJ{1}$ \\
\mr
{\UNEDFONEHFB}  & -45.600 & -143.935 & -234.380 & -260.437 & -73.946 & -51.913 \\
{\MAP}          & -46.157 & -139.972 & -245.287 & -250.964 & -77.353 & -70.143 \\
\br
\end{tabular}
\end{indented}
\end{table}
\begin{figure}[!ht]
\begin{center}
\includegraphics[width=0.9\linewidth]{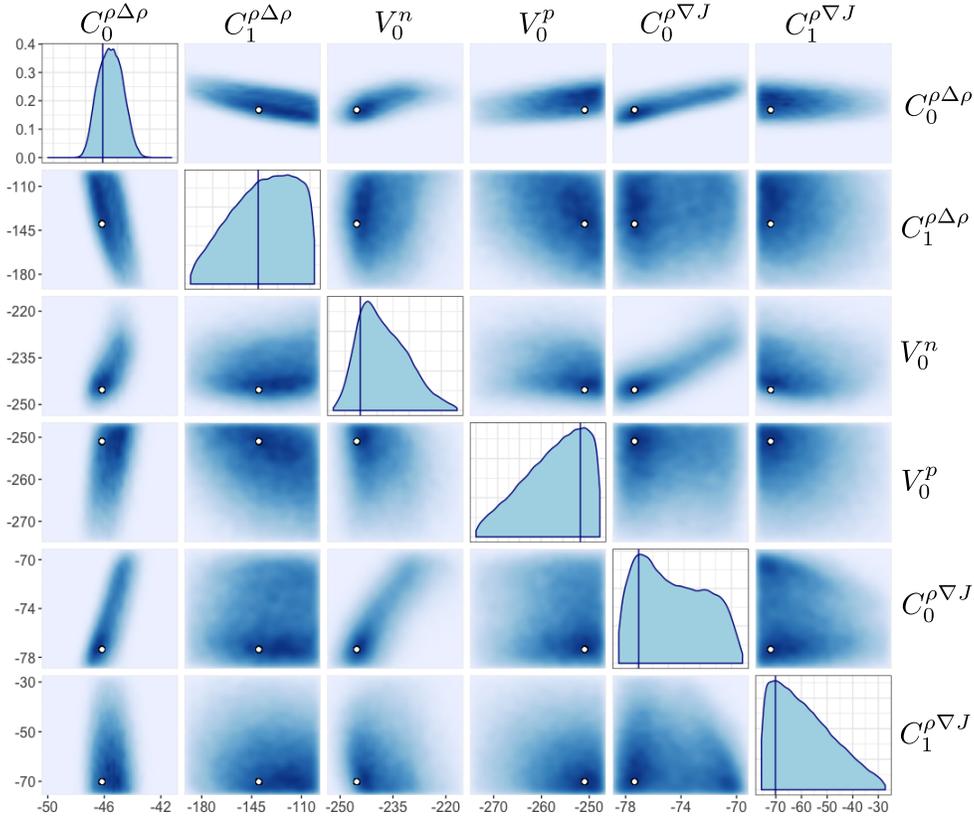}
\caption{Posterior estimates of the parameters. The blue lines on the diagonal 
and the white dots with black outlines on the off diagonals indicate the {\MAP} 
values.}
\label{fig:posteriorplot}
\end{center}
\end{figure}

Given the $\xb^{*}$ parameterization for the {\MAP} estimate, we can then run 
the {\GASP} emulator for this set of coupling constants and reconstruct the 
entire potential energy curve. The results are shown in \fref{fig:MAP} together 
with the curve obtained from the initial {\UNEDFONEHFB} parameterization. The 
shaded band around the {\MAP} curve shows the 95\% uncertainty from the {\GASP} 
emulator. Even if we only calibrated 3 excitation energies, {\EFI}, {\EA} and 
{\EB}, the Bayesian regression modifies the entire deformation energy curve. 
This could have a significant effect on observables such as spontaneous fission 
half lives $\tau_{\rm SF}$ which are related to the quantum-mechanical 
tunneling probability through the barrier and thus depend both on the barrier 
height but also its width. Qualitatively, the $\tau_{\rm SF}$ for $^{240}$Pu 
computed from the {\MAP} estimate could thus be a few orders of magnitude 
smaller than the value obtained with the initial parameterization of the 
functional, since the barriers are both lower and narrower.

\begin{figure}[!ht]
\begin{center}
\includegraphics[width=0.85\linewidth]{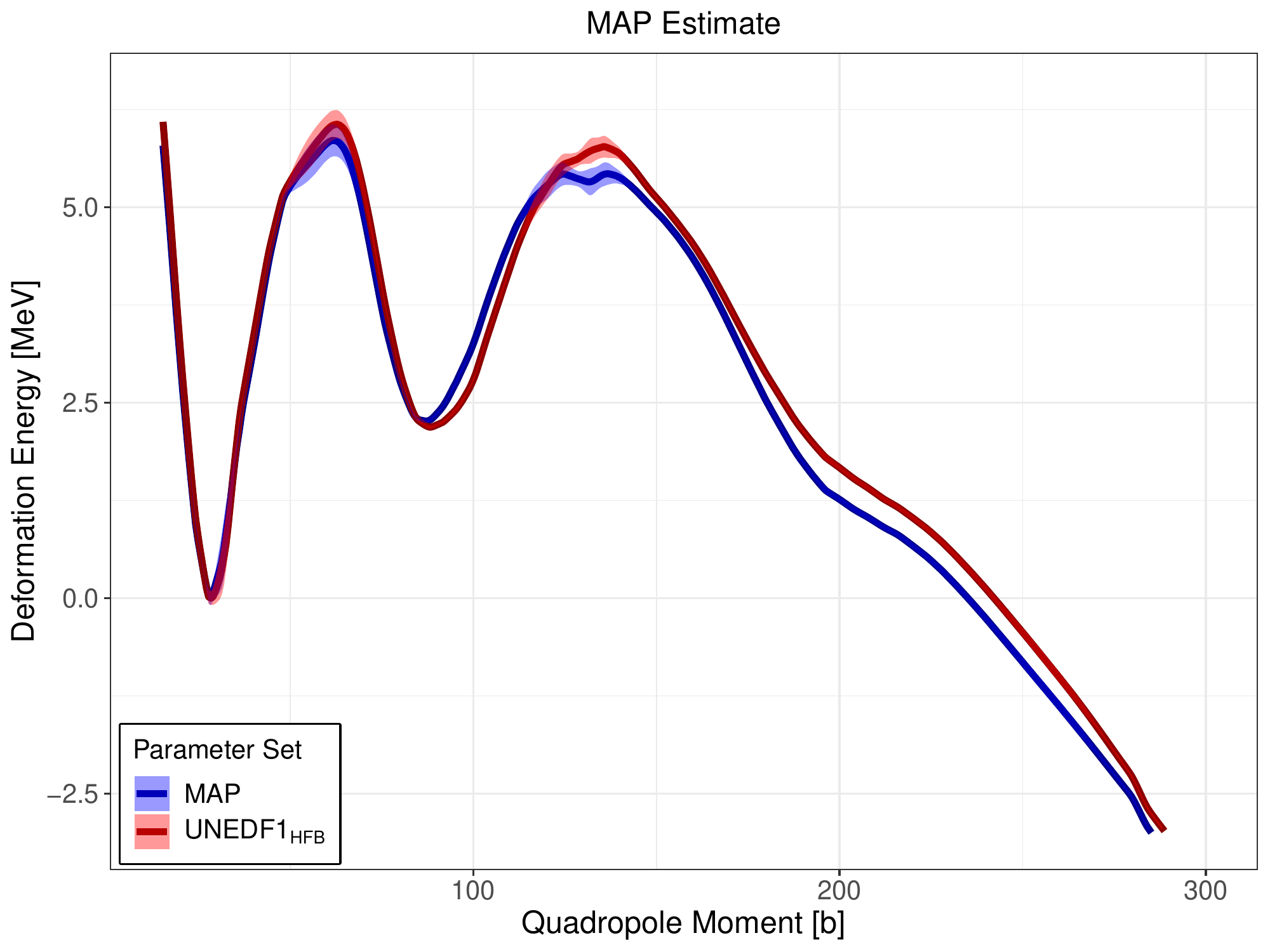}
\caption{Potential energy curve generated using the {\MAP} and {\UNEDFONEHFB} 
parameterizations. In both cases, the curve was obtained from the {\GASP} 
emulator, which provides the 95\% confidence region.}
\label{fig:MAP}
\end{center}
\end{figure}

The excitation energy and fission barrier heights can be extracted easily from 
the potential energy curves and are listed in \tref{tab:barriers}. The 
uncertainties are from the emulator uncertainty at the ground state and the 
relevant location for each energy. We recall that the empirical value used to 
determine the {\MAP} solution is a weighted average of three different numbers; 
see \tref{tab:exp_barriers}. In the case of the second barrier, one of the 
numbers is significantly different from the others, which leads to a larger 
uncertainty and explains why the {\MAP} result is further away from the data.

\Table{\label{tab:barriers} Energy of the fission isomer ($E_{\rm FI}$), 
height of the first ($E_{A}$) and second ($E_{B}$) barrier in MeV for the 
original {\UNEDFONEHFB} functional, the parameterization that maximizes the 
likelihood and empirical data. We also indicate the experimental error 
($\sigma_{\rm exp}$) that was used in the regression.}
\br
& {\UNEDFONEHFB} & {\MAP} & $\sigma_{\MAP}$ & Empirical & $\sigma_{\rm exp}$ \\
\mr
$E_{\rm FI}$  & 2.29 & $2.26$ & 0.048  & 2.35  & 0.20 \\
$E_{\rm A}$   & 6.15 & $5.85$ & 0.115  & 5.90 & 0.23 \\
$E_{\rm B}$   & 5.79 & $5.43$ & 0.085  & 5.28 & 0.31 \\
\br
\end{tabular}
\end{indented}
\end{table}

%%%%%%%%%%%%%%%%%%%%%%%%%%%%%%%%%%%%%%%%%%%%%%%%%%%%%%%%%%%%%%%%%%%%%%%%%%%%%%%
%%%%%%%%%%%%%%%%%%%%%%%%%%%%%%%%%%%%%%%%%%%%%%%%%%%%%%%%%%%%%%%%%%%%%%%%%%%%%%%
%%%%%%%%%%%%%%%%%%%%%%%%%%%%%%%%%%%%%%%%%%%%%%%%%%%%%%%%%%%%%%%%%%%%%%%%%%%%%%%
%%%%%%%%%%%%%%%%%%%%%%%%%%%%%%%%%%%%%%%%%%%%%%%%%%%%%%%%%%%%%%%%%%%%%%%%%%%%%%%

\section{Conclusions}
\label{sec:conclusions}

In this work, we investigated the use of Gaussian processes to quantify 
uncertainties in nuclear deformation properties. We built an emulator of the 
entire one-dimensional potential energy curve in $^{240}$Pu that is valid for a 
small 6-dimensional hypercube around the {\UNEDFONEHFB} parameterization of the 
Skyrme functional. When the potential energy curve follows a fully-connected 
path in the (infinite-dimensional) collective space, the numerical precision of 
the emulator is excellent (less than 100 keV); when ``discontinuities'' occur 
as, e.g., triggered by the onset of triaxiality, the quality of the emulation is 
a little degraded but remains rather good (less than 500 keV). The location of 
the scission point, which is marked as a discontinuity in the potential energy, 
is harder to pin-down, with uncertainties of the order of $\Delta q \approx \pm 
5$b. We used our emulator to determine the posterior distribution of the 
coupling constants of the Skyrme energy functional conditioned on the 
excitation energy of the fission isomer and the height of the two fission 
barriers. Most of the uncertainty comes from the empirical data.

Gaussian processes belong to the class of supervised learning techniques, which 
themselves are part of the broader field of machine learning. In nuclear 
density functional theory, these techniques can be applied on at least three 
different classes of ``data'': (1) the observable of interest, i.e., the total 
energy, radius, cross-section, etc. (2) matrix elements, e.g., of the mean 
field, pairing field or of some collective Hamiltonian and (3) the 
single-particle or quasiparticle wave functions. This work showed that there may 
be somewhat limited use in trying to emulate directly the observable of 
interest. The recent work of \cite{lasseri2020} suggests that working with 
matrix elements may hold more promise. Ultimately, one may have to directly 
work with individual wave functions themselves.

%%%%%%%%%%%%%%%%%%%%%%%%%%%%%%%%%%%%%%%%%%%%%%%%%%%%%%%%%%%%%%%%%%%%%%%%%%%%%%%
%%%%%%%%%%%%%%%%%%%%%%%%%%%%%%%%%%%%%%%%%%%%%%%%%%%%%%%%%%%%%%%%%%%%%%%%%%%%%%%
%%%%%%%%%%%%%%%%%%%%%%%%%%%%%%%%%%%%%%%%%%%%%%%%%%%%%%%%%%%%%%%%%%%%%%%%%%%%%%%
%%%%%%%%%%%%%%%%%%%%%%%%%%%%%%%%%%%%%%%%%%%%%%%%%%%%%%%%%%%%%%%%%%%%%%%%%%%%%%%

\section*{Acknowledgment}
This work was partly performed under the auspices of the U.S.\ Department of
Energy by Lawrence Livermore National Laboratory under Contract
DE-AC52-07NA27344 and was supported by the LLNL-LDRD Program under Project No. 
18-ERD-008. Computing support also came from the Lawrence Livermore 
National Laboratory (LLNL) Institutional Computing Grand Challenge program.

%%%%%%%%%%%%%%%%%%%%%%%%%%%%%%%%%%%%%%%%%%%%%%%%%%%%%%%%%%%%%%%%%%%%%%%%%%%%%%%
%%%%%%%%%%%%%%%%%%%%%%%%%%%%%%%%%%%%%%%%%%%%%%%%%%%%%%%%%%%%%%%%%%%%%%%%%%%%%%%
%%%%%%%%%%%%%%%%%%%%%%%%%%%%%%%%%%%%%%%%%%%%%%%%%%%%%%%%%%%%%%%%%%%%%%%%%%%%%%%
%%%%%%%%%%%%%%%%%%%%%%%%%%%%%%%%%%%%%%%%%%%%%%%%%%%%%%%%%%%%%%%%%%%%%%%%%%%%%%%

\section*{References}

\bibliographystyle{unsrt}
\bibliography{zotero_output,books,statisticspapers}

\end{document}